\documentclass[aps,prl,twocolumn,groupedaddress,showpacs]{revtex4}

\usepackage{graphicx}
\usepackage{dcolumn}
\usepackage{bm}
\usepackage{epsfig}
\usepackage{longtable}

\begin{document}


\title{Measurement of the Absolute np Scattering Differential Cross Section at 194 MeV}


\author{M. Sarsour$^{1}$}
\author{T. Peterson$^{1}$} \altaffiliation[Present address:
]{Dept. of Radiology and Radiological Sciences, Vanderbilt
University, Nashville, TN, USA}
\author{M. Planinic$^{1}$}
\altaffiliation[Present address: ]{Dept. of Physics, University of
Zagreb, Zagreb, Croatia}
\author{S.E. Vigdor$^{1}$}
\author{C. Allgower,$^{1}$}
\author{B. Bergenwall$^{2}$}
\author{J. Blomgren$^{2}$}
\author{T. Hossbach$^{1}$}
\author{W.W. Jacobs$^{1}$}
\author{C. Johansson$^{2}$}
\author{J. Klug$^{2}$}
\author{A.V. Klyachko$^{1}$}
\author{P. Nadel-Turonski$^{2}$}
\author{L. Nilsson$^{2}$}
\author{N. Olsson$^{2}$}
\author{S. Pomp$^{2}$}
\author{J. Rapaport$^{3}$}
\author{T. Rinckel$^{1}$}
\author{E.J. Stephenson$^{1}$}
\author{U. Tippawan$^{2,4}$}
\author{S.W. Wissink$^{1}$}
\author{Y. Zhou$^{1}$\\}
\affiliation{$^{1}$Indiana University Cyclotron Facility and
Department of Physics, Bloomington, IN 47408, USA\\
$^{2}$Uppsala University, Uppsala, Sweden\\
$^{3}$Ohio University, Athens, OH, USA\\
$^{4}$Chiang Mai University, Chiang Mai, Thailand\\ }


\date{\today}

\begin{abstract}
We describe a double-scattering experiment with a novel tagged
neutron beam to measure differential cross sections for np
back-scattering to better than $\pm 2\%$ absolute precision.  The
measurement focuses on angles and energies where the cross section
magnitude and angle-dependence constrain the charged pion-nucleon
coupling constant, but existing data show serious discrepancies
among themselves and with energy-dependent partial wave analyses
(PWA).  The present results are in good accord with the PWA, but
deviate systematically from other recent measurements.

\end{abstract}


\maketitle

\bigskip

The neutron-proton elastic scattering database at intermediate
energies is plagued by experimental inconsistencies and cross
section normalization difficulties \cite{Bon1978,Hur1980,Rah1998}.
These problems have led the most sophisticated partial wave
analyses (PWA) of the data \cite{Sto1993,Bug1995,Ren2001} to
ignore the majority (including the most recent) of measured cross
sections, while the literature is filled with heated debates over
experimental and theoretical methods \cite{Ren1998,Blo2000},
including proposed radical ``doctoring" (angle-dependent
renormalization) to ``salvage" allegedly flawed data
\cite{deS2002}. Meanwhile, an empirical evaluation of a
fundamental parameter of meson-exchange theories of the nuclear
force -- the charged $\pi$NN coupling constant, $f_c^{2}$ -- hangs
in the balance \cite{Blo2000}. We report here the results of a new
experiment, carried out utilizing quite different techniques from
earlier measurements in an attempt to resolve the most worrisome
experimental discrepancies.

The present experiment involves a kinematically complete
double-scattering measurement to produce and utilize a ``tagged"
intermediate-energy neutron beam \cite{Pet2004}, thus greatly
reducing the usual systematic uncertainties associated with the
determination of neutron flux.  Products from the second
scattering were detected over the full angle range of interest
simultaneously in a large-acceptance detector array, to eliminate
the need for cross-normalization of different regions of the
angular distribution.  The use of carefully matched solid CH$_2$
and C targets permitted frequent measurement and accurate
subtraction of quasifree scattering background, thereby minimizing
reliance on kinematic cuts to isolate the free np-scattering
sample.  These methods, combined with multiple internal
crosschecks built into the data analysis procedures, have allowed
us to achieve systematic error levels in the absolute cross
section below 2\%.  In addition to addressing the previous
discrepancies, the results provide a useful absolute cross section
calibration for intermediate-energy neutron-induced reactions.

\begin{figure}[htp!]
\includegraphics[scale=0.32]{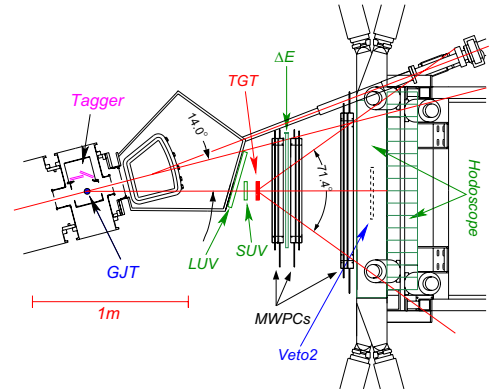}
\caption{\label{fig:setup} Top view of the np scattering
experiment setup.}
\end{figure}

The experiment was carried out during the final year of operation
of the Indiana University Cyclotron Facility's Cooler ring
\cite{Pol1991}, with apparatus illustrated in Fig.~\ref{fig:setup}
and described in detail in \cite{Pet2004}. Neutrons of 185-197 MeV
were produced via the charge-exchange reaction p+d $\rightarrow$
n+2p, initiated by a stored, electron-cooled 203 MeV proton beam,
with typical circulating current of 1-2 mA, in a windowless
deuterium gas jet target (GJT) of thickness $\approx 2 - 4 \times
10^{15}$ atoms/cm$^{2}$. The ultra-thin target permitted detection
of the two low-energy recoil protons in an array (``tagger" in
Fig.~\ref{fig:setup}) of four 6.4$\times$6.4 cm$^2$ double-sided
silicon strip detectors (DSSD) with 480 $\mu$m readout pitch in
two orthogonal directions, each followed by a silicon pad
(``backing") detector (BD) of the same area. Only recoil protons
($\lesssim 11$ MeV) that stopped in either the DSSD's or BD's were
considered in the data analysis. Measurements of energy, arrival
time, and two-dimensional position for both recoil protons in the
tagger, when combined with precise knowledge of cooled p beam
direction and energy, allowed 4-momentum determination for each
produced neutron on an event-by-event basis. The (uncollimated)
tagged neutrons were distributed over a significant range in
energy and angle, but these parameters were measured for each
produced neutron with resolutions $\sigma_E \approx 60$ keV and
$\sigma_{angle} \approx$ 2 mrad.

The forward setup included a solid secondary scattering target of
CH$_{2}$ or graphite positioned 1.1 m downstream of the GJT,
centered on a neutron production lab angle of 14.0$^\circ$. Both
solid targets had transverse dimensions 20$\times$20 cm$^2$ and
thickness of $0.99 \times 10^{23}$ carbon atoms/cm$^2$. The
typical flux of tagged neutrons with energy above 185 MeV
intercepting the secondary target was 200 Hz.  Two upstream
plastic scintillators (LUV and SUV in Fig.~\ref{fig:setup}) vetoed
tagged neutrons that interacted before the secondary target.
Following the target was a forward array of plastic scintillators
for triggering and energy information and a set of three (3-plane)
multi-wire proportional chambers (MWPCs) to track forward protons.
The forward detector acceptance was nearly 100\% for np scattering
events with $\theta_{c.m.} \geq 130^\circ$, falling to 50\% by
$\theta_{c.m.} = 90^\circ$. The MWPC between the secondary target
and $\Delta$E scintillator allowed discrimination against np
events initiated in that scintillator. The rear hodoscope
comprised 20 plastic scintillator bars of sufficient thickness (20
cm) to stop 200 MeV protons and give 15-20$\%$ detection
efficiency for 100-200 MeV neutrons.

Specially designed DSSD front-end electronics \cite{Pet2004}
permitted a tagger-based event trigger on neutron candidates
(consistent with two distinct tagger hits and no accompanying
signals from LUV or SUV), whether or not the neutrons interacted
in the forward target and/or detectors. Tagged neutron events were
recorded in three mutually exclusive event streams \cite{Pet2004},
coupling the tagger trigger with (1) no rear hodoscope coincidence
(providing a prescaled sample for neutron flux monitoring); or a
coincidence with (2) both $\Delta$E scintillator and rear
hodoscope (for np scattering candidates); or (3) rear hodoscope
but not $\Delta$E (for evaluating the neutron detection efficiency
of the hodoscope). Comparative analyses of the three separate
event streams, with respective yields $N_1, ~N_2$ and $N_3$,
facilitated crosschecks to calibrate the system \cite{Pet2004} and
to study potential systematic errors.

Neutron beam properties were defined by identical cuts for all
three event streams, so that associated systematic uncertainties
would cancel in the yield ratios from which the absolute np
scattering cross section is extracted.  Among the common cuts are
ones on DSSD vs.\ BD energy deposition in the tagger
\cite{Pet2004}, used to select two tagged neutron classes for
analysis: (a) ``2-stop" events, where both recoil protons stopped
inside the DSSD's (either the same or different quadrants of the
tagger); (b) ``1-punch" events, where one of two recoil protons
incident on different quadrants punched through into the
corresponding BD and stopped there. These classes differ
significantly in neutron energy ($E_n$) and position profiles
\cite{Pet2004}, allowing an important crosscheck on the accuracy
of the tagging technique by comparing np cross sections extracted
independently from each class. Other common cuts defined a
fiducial area for neutrons impinging on the secondary target and
eliminated common-mode BD noise (via pulse height correlations
among quadrants) that sometimes led to misidentification of event
class.

Additional misidentification discovered during data analysis was
attributed to an electronics malfunction in the gating or clearing
circuit for one analog-to-digital converter module, that removed
all valid BD energy signals for some fraction of events. The
corrupted events were misidentified as 2-stop events, with
systematically incorrect predictions of tagged neutron trajectory
(since some recoil proton energy was missed), and hence of np
scattering angle for event stream 2. A subsample of these
corrupted events could be isolated by means of their valid BD
timing signals, and their properties were accurately reproduced by
appropriately scaling the surviving sample of all events with
valid BD energies and times, after setting these energies to zero
in software. Thus, the surviving punch-through events permitted a
reliably unbiased subtraction of the corrupted 2-stop events,
independently for each event stream. The subtraction confirmed
that the same fraction (typically 23\%) of punch-through events
was lost from each event stream, with no net effect on the
extracted 1-punch cross sections.

Kinematic cuts applied exclusively to event stream 2 to define np
free-scattering events from the secondary target were used
sparingly. We relied instead on accurate background subtraction
facilitated by frequent interchange of the carefully matched
CH$_{2}$ and C targets. The CH$_2$ and C runs were normalized via
the pd elastic scattering yield from the GJT measured in a fourth
event stream for the two targets. The pd events were identified by
their clear kinematic locus in the energy of recoil deuterons
detected in the tagger \emph{vs.}\ the position of coincident
forward protons in the front MWPC. The subtraction removed not
only quasifree scattering off carbon nuclei in the target, but
also background from other sources, such as tagged n scattering
from the aluminum support platform on which the secondary target
sat, or protons produced in the GJT that passed above the top of
the LUV and SUV scintillators, mocking up np back-scattering
events.

\begin{figure*}[htb]
\includegraphics[scale=0.18]{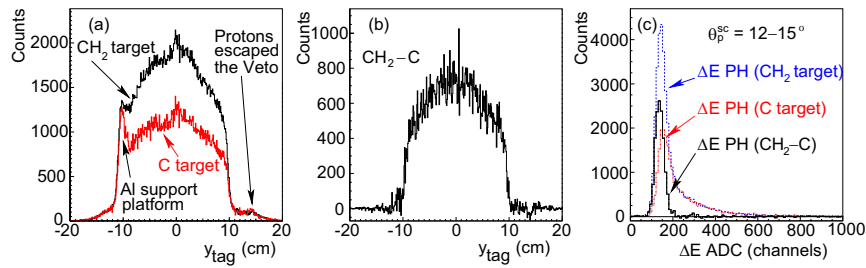}
\caption{\label{fig:bkgrnd} Distributions of np scattering
candidate events in $y_{tag}$ (a,b) and $\Delta$E (c) for CH$_{2}$
and C targets and their difference.}
\end{figure*}

The success of the background subtraction is illustrated in
Fig.~\ref{fig:bkgrnd}, where (a) and (b) show the vertical
position ($y_{tag}$) of neutrons on the secondary target, as
reconstructed from the tagger, for CH$_{2}$ and normalized C
targets, and for their difference. Figure~\ref{fig:bkgrnd}(c)
shows the $\Delta$E pulse-height spectrum within a given
scattering angle ($\theta_p^{sc}$) bin for both targets and for
their difference. Prominent background features associated both
with the secondary target (\emph{e.g.}, the long quasifree
scattering tail in $\Delta$E) and with other sources (\emph{e.g.},
the peak in frame (a) from the Al support platform) are
simultaneously accurately removed by the subtraction.  The
subtraction reveals in frame (b) a $y_{tag}$ distribution
reflecting the tagged n (2-stop + 1-punch) beam profile,
convoluted with the np scattering cross section, forward detector
acceptance and sharp CH$_2$ target edges (the sharpness
illustrating the good spatial resolution of the tagging).

Background-subtracted spectra such as that in
Fig.~\ref{fig:bkgrnd}(c) were used to evaluate efficiencies for
the few loose cuts imposed on event stream 2 to improve the
free-scattering signal-to-background ratio, including ones on
$\Delta E$ {\em vs.} $\theta_p^{sc}$ and on MWPC proton track
quality. Cuts on the hodoscope pulse height were avoided, to
remove reliance on detailed understanding of the nuclear reaction
tail for protons stopping in this thick scintillator.

\begin{figure}[htb]
\includegraphics[scale=0.40]{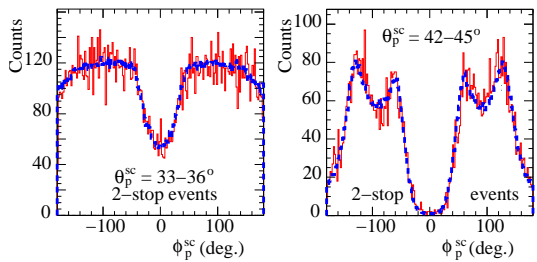}
\caption{\label{fig:accep} The distribution of free (CH$_2$-C) np
scattering events with respect to proton azimuthal angle $\phi$,
for two lab scattering angle bins. The solid (dashed) histograms
are measured (simulated and normalized to fit the measurements).}
\end{figure}

The forward detector acceptance was determined as a function of
$\theta_p^{sc}$ from simulations matched to \emph{measured}
distributions of np free-scattering events in proton azimuthal
angle $\phi_p$.  In the simulations, the longitudinal coordinate
of the n production vertex within the GJT and its transverse
coordinates on the secondary target were generated randomly for
each event, but within distributions determined from the
experiment.  These coordinates determined the incident n angle.
Generated outgoing p trajectories were accepted if they would
yield signals above the hodoscope threshold (required in trigger)
and in all three MWPCs (required in the data analysis). Forward
detector location parameters were tuned to reproduce the measured
$\phi_p$ distributions for all $\theta_p^{sc}$ bins and for
1-punch and 2-stop samples simultaneously. Typical fits in
Fig.~\ref{fig:accep} reveal structure of purely geometric origin,
from rectangular detector edges projected on $\theta$ and $\phi$.
For $\theta_p^{sc} \leq 24^\circ$ the measured and simulated
$\phi$ distributions are uniform, since the scattered protons are
completely contained within the forward array.

Absolute differential cross sections were obtained from the yields
in event streams 1, 2 and 3 defined above:
\begin{equation}
\biggl( {d \sigma \over d
\Omega}\biggr)_{lab}=\frac{N_2(\theta_p^{sc}) \prod c_i}{(N_1 +
N_2 + N_3) t_H |d \cos(\theta_{p}^{sc})| a_{\phi}(\theta_p^{sc})},
\end{equation}
where $N_j$ represents the number of events (corrected for
prescaling where appropriate) surviving all relevant cuts and
background subtractions for event stream $j$; the $c_i$ are small
corrections, summarized in Table~\ref{tab:syserr}, for
inefficiencies, tagged neutron losses or backgrounds, and software
cut and dead time differences among event streams; $t_H = (1.988
\pm 0.008) \times 10^{23}$ H atoms/cm$^2$ for the CH$_2$ target;
and $a_{\phi}$ is the azimuthal acceptance determined from
simulations for the given angle bin.  The data were analyzed in 1
MeV wide $E_n$ slices from 185 to 197 MeV, and among the $c_i$ for
each slice are small (always $<1\%$) corrections, based on the
theoretical energy dependence calculated with the Nijmegen PWA
\cite{Sto1993}, to extract an effective cross section at $\langle
E_n \rangle =194.0 \pm 0.15$ MeV.

\begin{figure}[htp!]
\includegraphics[scale=0.30]{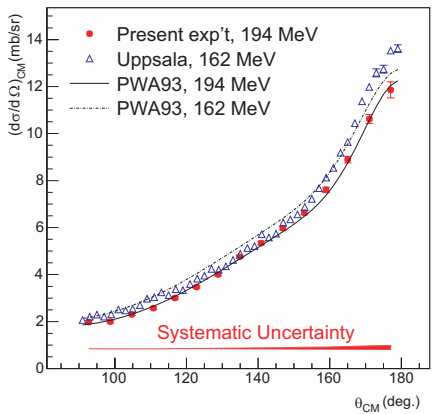}
\caption{\label{fig:avexs} Absolute differential cross section
from the present experiment, compared with data from
Ref.~\cite{Rah1998} and with PWA calculations at two relevant
energies.  The error bars and shaded band for the present results
represent, respectively, statistical (including background
subtraction) and systematic (including overall normalization)
uncertainties.}
\end{figure}

Cross sections extracted independently for the 1-punch and 2-stop
samples agree within statistical uncertainties
($\chi^2$/point$\approx$1.0) in both magnitude and angular shape.
This comparison supports the reliability of the experiment and
analysis, as these events come from complementary regions of the
tagged beam spatial and energy profiles \cite{Pet2004}. Cross
sections extracted for different time periods within the
production runs, and with different sets of cuts, are also
consistent within uncertainties. The results, averaged over the
2-stop and 1-punch samples, are compared in Fig.~\ref{fig:avexs}
with previous experimental results at 162 MeV \cite{Rah1998} and
with the Nijmegen partial wave analysis (PWA93) at the two
relevant energies \cite{nij}.

By using a tagged beam, we have sacrificed statistical precision
for better control of systematic errors, in order to assess which
previous experiments may have suffered from unrecognized
systematic problems. Each systematic uncertainty summarized in
Table~\ref{tab:syserr} has been evaluated in a separate analysis,
sometimes involving auxiliary measurements. The errors are, except
where noted otherwise in the Table, angle-independent
normalization uncertainties.  The largest correction to the data
and attendant uncertainty arise from cuts to remove stream 2 (but
not 1 or 3) events where the np scattering vertex transverse
coordinates predicted from n tagging \emph{vs.} p ray-tracing
disagree by more than three times the resolution $\sigma$.  The
removed events (6.3\% of the total sample) are affected by several
factors -- \emph{e.g.}, sequential reactions in the secondary
target and upstream material, or errors associated with recoil
protons stopping in dead layers within the tagger -- that lead to
ambiguities in neutron energy and scattering angle. Combining all
effects in Table~\ref{tab:syserr} and summing uncertainties in
quadrature, the net correction ($\prod c_i$) applied to raw cross
sections is $\approx 1.10 \pm 0.015$, with an angle-dependence of
the systematic uncertainty indicated by the shaded band in
Fig.~\ref{fig:avexs}.

The present results are in quite good absolute agreement with the
Nijmegen PWA93 calculations, over the full angular range covered.
The small deviations seen might be removed by minor tuning of
phase shifts. In contrast, the present results deviate
systematically, especially in the steepness of the back-angle
cross section rise, from earlier measurements
\cite{Rah1998,Hur1980} that the Nijmegen group had rejected in
their analyses by applying controversial criteria. These
deviations are larger than the differences expected from the
neutron energy changes among the various experiments.  As the
back-angle rise is particularly influential in pole extrapolations
used \cite{Eri1996,Blo2000} to extract the pion-nucleon-nucleon
coupling constant, the present data strongly favor the value
($f_c^2 = 0.0748 \pm 0.0003$) given by the Nijmegen \cite{Ren2001}
and other \cite{Bug1995} partial-wave analyses.

\begin{table}[htb]
\caption{\label{tab:syserr} Correction factors ($c_i$) and
systematic uncertainties in correction factors for the np cross
sections.}
\begin{ruledtabular}
\begin{tabular}{lll}
Source & Correction Factor & Uncertainty \\
\hline\hline
Accid. tagger coinc.                  & 1.0003               & $<$ $\pm$ 0.001 \\
Non-D$_{2}$ tagger                  & 1.0067 (2-stop);     & $\pm$ 0.002 \\
~~background                        &  1.0044 (1-punch)    & \\
n pos'n unc. on CH$_{2}$             & 1.0000               & $\pm$ 0.001 \\
n atten'n before CH$_{2}$                  & 1.005                & $\pm$ 0.0025 \\
C bkgd. subtraction                       & 1.0000               & $\pm$ 0.004 \\
Reaction tail losses                           & 1.004                & $\pm$ 0.002 \\
Neutron polarization                    & Angle-dependent:    & $\pm$ 0.001 \\
~~effects                               & $>0.9988$ (1-punch)    & \\
                                        & $<1.0014$ (2-stop)     & \\
Software cut losses                       & 1.010                & $\pm$ 0.005 \\
Sequential react'ns                       & 1.063    & $\pm$ 0.010 \\
~~\& $x_{tag}$(n) errors                  & & \\
CH$_{2}$ tgt. thickness                      & 1.0000               & $\pm$ 0.004 \\
np scattering                  & 1.0000 & $\leq\pm 0.001$ ($>$120$^\circ$)  \\
~~acceptance                         &  & $\rightarrow \pm 0.017$ (90$^\circ$)\\
MWPC efficiency                                & 1.017                & $\pm$0.002 \\
Trigger inefficiency                           & 1.002 + 0.008 $\times$ & $\pm$ [0.001 + 0.004 \\
                                               & cos$^{2}$($\theta_{p}^{LAB}$)
                                               & $\times$ cos$^{2}$($\theta_{p}^{LAB}$)]\\
Dead time diffs.                          & 0.991                & $\pm$ 0.005 \\
Scattering angle                        & 1.000                & angle-dependent, \\
~~errors                                &                      & ~~$\leq \pm$0.004 \\
Corruption subt'n                        & 1.000                & $< \pm 0.001$ \\
\hline {\bf Net}                    & $\approx 1.10$       & $\approx \pm 0.015$ \\
\end{tabular}
\end{ruledtabular}
\end{table}

\begin{acknowledgments}
We thank the operations staff of the Indiana University Cyclotron
Facility for providing the superior quality cooled beams, and Hal
Spinka and Catherine Lechanoine-Leluc for the loan of critical
detector hardware, needed for successful execution of this
experiment. We acknowledge the U.S. National Science Foundation's
support under grant numbers NSF-PHY-9602872 and 0100348.
\end{acknowledgments}


\end{document}